\documentclass[preprintnumbers, floatfix, letterpaper, twocolumn,aps,prd,epsfig,nofootinbib,natbib,longbibliography]{revtex4-2}
\usepackage{graphicx}
\usepackage{epstopdf}
\usepackage{latexsym}
\usepackage{amssymb}
\usepackage{amsmath}
\usepackage{color}
\usepackage{mathrsfs}
\usepackage{xparse}
\usepackage{bbding}
\usepackage{pifont}
\usepackage{comment}
\usepackage{ulem}
\usepackage{float}
\usepackage[inline]{enumitem}
\usepackage{soul}
\usepackage{scalerel}
\usepackage[caption=false]{subfig}

\let\Right\right
\let\Left\left
\makeatletter
\def\right#1{\Right#1\@ifnextchar){\!\right}{}}
\def\left#1{\Left#1\@ifnextchar({\!\left}{}}
\makeatother

\usepackage[
            pdfstartview=FitH,
            bookmarksnumbered=true,
            bookmarksopen=true,
            colorlinks,
            linkcolor=blue,
            anchorcolor=green,
            citecolor=blue
            ]{hyperref}
\begin{document}

%%%%%%%%%%%%%%%%%%%%%%%%%%%%%%%%%%%%%%%%%%%%%%%%%%%%%%%%%%%%%%%
\renewcommand\arraystretch{2}
\newcommand{\bq}{\begin{equation}}
\newcommand{\eq}{\end{equation}}
\newcommand{\bqn}{\begin{eqnarray}}
\newcommand{\eqn}{\end{eqnarray}}
\newcommand{\nb}{\nonumber}
\newcommand{\lb}{\label}

\newcommand{\La}{\Lambda}
\newcommand{\va}{\scriptscriptstyle}
\newcommand{\be}{\nopagebreak[3]\begin{equation}}
\newcommand{\ee}{\end{equation}}

\newcommand{\ba}{\nopagebreak[3]\begin{eqnarray}}
\newcommand{\ea}{\end{eqnarray}}

\newcommand{\la}{\label}
\newcommand{\n}{\nonumber}
\newcommand{\su}{\mathfrak{su}}
\newcommand{\SU}{\mathrm{SU}}
\newcommand{\U}{\mathrm{U}}
\newcommand{\red}{ }

\newcommand{\R}{\mathbb{R}}

\newcommand{\cb}{\color{blue}}
\newcommand{\cc}{\color{cyan}}
\newcommand{\cm}{\color{magenta}}
\newcommand{\rc}{\rho^{\scriptscriptstyle{\mathrm{I}}}_c}
\newcommand{\rd}{\rho^{\scriptscriptstyle{\mathrm{II}}}_c}
\NewDocumentCommand{\evalat}{sO{\big}mm}{%
\IfBooleanTF{#1}
{\mleft. #3 \mright|_{#4}}
{#3#2|_{#4}}%
}
\newcommand{\PRL}{Phys. Rev. Lett.}
\newcommand{\PL}{Phys. Lett.}
\newcommand{\PR}{Phys. Rev.}
\newcommand{\CQG}{Class. Quantum Grav.}
%%%%%%%%%%%%%%%%%%%%%%%%%%%%%%%%%%%%%%%%%%%%%%%%%%%%%%%%%%%%%%%

%%%%%%%%%%%%%%%%%%%%%%%%%%%%%%%%%%%%%%%%%%%%%%%%%%%%%%%%%%%%%%%

\title{Analytical models of supermassive black holes in galaxies surrounded by dark matter halos}

\author{Zibo Shen${}^{a,b}$}
\email{Zibo$\_$Shen@baylor.edu}

\author{Anzhong Wang${}^{a}$ \footnote{Corresponding author}}
\email{anzhong$\_$wang@baylor.edu; Corresponding author}

\author{Yungui Gong${}^{c}$}
\email{gongyungui@nbu.edu.cn}

\author{Shaoyu Yin${}^{d}$}
\email{syyin@zjut.edu.cn}

\affiliation{${}^{a}$ GCAP-CASPER, Physics Department, Baylor University, Waco, Texas 76798-7316, USA\\
	${}^{b}$ Department of Astronomy,
	School of Astronomy and Space Science, University of Science and Technology of China, Hefei, Anhui 230026, China\\
	${}^{c}$Department of Physics, School of Physical Science and Technology, Ningbo University, Ningbo, Zhejiang 315211, China\\
	${}^{d}$ Institute for Theoretical Physics and Cosmology, Zhejiang University of Technology,
	Hangzhou 310023, China.}

\date{\today}

\begin{abstract}

In this Letter, we present five analytical models in closed forms, each representing a supermassive black hole (SMBH) located at the center of a galaxy surrounded by dark matter (DM) halo. The density profile of the halo vanishes inside twice the Schwarzschild radius of the hole and satisfies the weak, strong, and dominant energy conditions. The spacetime are asymptotically flat, and the difference among the models lies in the slopes of the density profiles in the spike and regions far from the center of the galaxy. Three of them represent cusp models, whereas the other two represent core models.  {With the well-known (generalized) Newman-Janis algorithm, rotating
SMBHs with DM halos can be easily constructed from these models.}

\end{abstract}

\maketitle

\section{
 Introduction
}

The overwhelming evidence for dark matter (DM) is clear and compelling on a wide range of scales  \cite{BHS05,Freese09,SBD17,WT18,AM21}.
Motion of galaxies within clusters  \cite{Zwicky33,Zwicky37},  galaxy rotation curves \cite{RF70,Freeman70},  bullet cluster systems \cite{Clowe06},
baryonic acoustic oscillation  \cite{Eisenstein05} and  cosmic microwave background  \cite{WMAP09} all point to its existence.
According to the latest Planck  data \cite{Planck18}, about $5/6$ of the mass in the Universe is made of DM.    However, despite numerous efforts in the search for DM in the past decades, our understanding is still based only on their gravitational interactions. Thus, uncovering the nature and origin of DM is arguably one of the greatest challenges of modern physics and cosmology \cite{Cebrain23,Smarra23,MR23}.

According to the galaxy formation scenario, every galaxy forms within a DM halo, and at the center of almost every galaxy a supermassive black hole (SMBH) is present \cite{VME04}. In fact, two such SMBHs have already been observed by the Event Horizon Telescope (EHT),
one is the  Sagittarius $A^*$   black hole (BH) at the center of our own Milky Way galaxy \cite{EHT22},
and the other called $M87^*$ is  at the center of the more distant Messier 87 galaxy \cite{EHT19}.

With the arrival of the gravitational wave
(GW)  astronomy \cite{GWTC-3}, the studies of SMBHs in galaxies with DM halos have gained further momentum (See, for example,
\cite{ERS20,Padilla21,Marasco21,Bansal23,Zhang23}
and references therein). This is because such studies have important consequences for compact objects orbiting around a SMBH.
Surrounding such SMBHs to distances of a few parsecs, a dense cusp of millions of compact objects is present \cite{Schodel14}. These include stellar-mass black holes, neutron stars, and white dwarfs, some of which move along orbits around the SMBH and form two-body gravitational systems with the SMBH. These systems give rise to either intermediate mass-ratio spirals (IMRIs) or extreme mass-ratio spirals (EMRIs), depending on the mass of the SMBH.
In either case, the GWs emitted by such systems will be the primary sources for the forthcoming generation of GW detectors. In particular, the mass ratio $q \equiv m/M$ of such binaries is typically of ${\cal{O}}(10^{-3} \sim 10^{-4})$ for IMRIs \cite{Amaro-Seoane07}, and ${\cal{O}}(10^{-5} \sim 10^{-6})$ for EMRIs \cite{Babak17}.
GWs emitted by IMRIs can be detected by upcoming ground-based detectors, such as the Einstein Telescope \cite{ET} and the
Cosmic Explorer \cite{CE},  as well as space-based ones, such as LISA \cite{LISA}, TianQin \cite{TianQin},
Taiji \cite{Taiji} and DESIGO \cite{ DECIGO}, while the ones from EMRIs are exclusively for space-based detectors.

When a compact object moves inside the DM halo, gravitational drag is induced on the inspiring object, changing its orbit and the gravitational waveforms emitted by it.
For IMRIs, such effects can be detected using LISA
\cite{Eda15,Kavanagh20,Coogan22,K3Y23}. Similar effects for EMRIs are also expected and can be used to study the nature of  DM \cite{Hannuksela20,Zi21,Fan22,Berezhiani23}.
For binaries with a very small $q$, to the leading order, the compact object can be well described as a test particle, moving under the influence of the SMBH and DM halo \cite{Cardoso22a}.
Therefore, in this Letter, we focus on the gravitational field produced by an SMBH in a galaxy with a DM halo.

Using Newtonian gravity, Gondolo and Silk first studied the effects of a SMBH sitting at the center of a galaxy on the distribution of the DM halo and found that a spike can be produced at a distance of $4 r_s$,
where $r_s$ denotes the Schwarzschild radius of the SMBH \cite{GS99}. The orbits of the DM particles with radii less than  $4 r_s$ become unstable, so the density of the DM halo vanishes for  $r \lesssim 4 r_s$. Sadeghian, Ferrer, and Will later considered the relativistic effects and found that a spike is indeed developed \cite{SFW13}. However, they found that the orbits of the DM particles now start to be unstable at $2r_s$, instead of $4 r_s$, purely due to relativistic effects, which is further confirmed by Speeney {\it et al} \cite{SABB22} (See also \cite{DeLuca23}). Due to the complexity of the problem, such studies have been carried out mainly numerically, except for one particular case \cite{Cardoso22}, in which Cardoso {\it et al} constructed an analytic model with a density profile that asymptotically approaches the Hernquist density contribution \cite{Hernquist90}. This solution has immediately attracted much attention, and its various properties and applications have already been explored, including its Love numbers, quasi-normal modes (QNMs), tails and gray-body factors \cite{Cardoso22,Konoplya21,Liu22,Desrounis23}, shadows  \cite{Xavier23}, and the detectability of GWs emitted by am EMRI by LISA, TianQin and Taiji \cite{Dai23,Rahman23}, to name a few of them. Despite all these interesting properties and aspects, the model suffers from a couple of defects. For example, the density of the DM halo vanishes only inside the BH horizon, instead of vanishing in the region $r \le 2r_s$, as shown explicitly in \cite{SFW13,SABB22}. In addition, near the BH horizon, the DM halo does not satisfy the dominant energy condition \cite{Hawking:1973uf}.
Extensions of this model to different circumstances have also been studied; see, for example,  \cite{Nampalliwar21,KZ22,Liu23}  and references therein.

In this Letter, we present five analytic solutions, each representing a SMBH located at the center of a galaxy surrounded by a DM halo, in which the energy density of the halo satisfies all the energy conditions, weak, strong and dominant \cite{Hawking:1973uf}, and identically vanishes for $r \le 2r_s$. The spacetime in the region $r \lesssim 2r_s$ is described by the vacuum Schwarzschild BH solution, which is smoothly matched to the exterior across $r = 2r_s$. The difference between these models lies in the slopes of the density profiles in the spike and far regions. To the best of our knowledge, these are the only analytic models found so far that have such desirable properties.

The detailed derivations of these solutions and their main properties are presented in  \cite{Shen23}.  { In addition, we also find other analytical solutions with different combinations of the four parameters
$(n, \alpha, \beta, \gamma)$  introduced in (\ref{eq2}). However, their derivations and expressions are much more complicated than those presented here, so we shall report them on another occasion.}

\section{
General Setup of the Models
}

Let us consider an SMBH sitting at the center of a galaxy and surrounded by a cloud made of massive DM particles. Cosmologically, DM is described by a perfect fluid with vanishing pressures. However, such a cloud cannot exist within spherically symmetric spacetime because the attractive forces from the SMBH will always lead the cloud to collapse into the BH. To capture the main properties of the realistic situation, here we consider a spherical Einstein cloud made of collisionless DM particles all moving tangentially with zero total angular momentum, so that the cloud will have a negligible pressure only along the radial direction, while the tangential pressures prevent the DM particles from falling into BHs \cite{Cardoso22}. For such a cloud, the corresponding energy-momentum tensor of the DM halo is given by $T^{\mu}_{\nu} = (-\rho(r), 0, P(r), P(r))$, here $\rho$ and $P$ denote respectively  the energy density and tangential pressures of the DM halo. The spacetime are describe by  $ds^2 = - f(r) dt^2 + (1- 2m(r)/r)^{-1}dr^2  + r^2 (d\theta^2 + \sin^2\theta d\phi^2)$,
where  $f(r)$ and  $m(r)$ are the solutions of the Einstein field equations,  {$G^{\mu}_{\nu} = 8\pi T^{\mu}_{\nu}$. Note that in this letter we choose units such that $c = G = 1$, where $c$ and $G$ denote the speed of light and the
Newtonian constant, respectively.}
Then, it can be shown that the field equations have three independent components, given respectively by
\bqn
\lb{eq1}
\frac{f'(r)}{f(r)}  &=& \frac{{2m(r)}}{r\left(r - 2m(r)\right)},  \\
\lb{eq1b}
m'(r)  &=&  4\pi   r^2 \rho(r),\\
\lb{eq1c}
P(r) &=&    \frac{m(r) \rho(r)}{2\left(r - 2m(r)\right)},
\eqn where a prime denotes the derivative wrt $r$.  To uniquely determine the four unknown functions $(f, m, \rho, P)$, in this Letter, we consider the choice \cite{Hernquist90,Zhao96,SABB22,FMC23}
\bq
\lb{eq2}
\rho(r)   = \frac{\rho_0\left(1- {4M }/{r}\right)^n}{({r}/{a})^{\gamma}\left(1+ ({r}/{a})^{\alpha}\right)^{(\beta - \gamma)/\alpha}},
\eq where $\rho_0$ is the characteristic density of the halo,  $M $  the BH mass, $a$ is the characteristic scale of the DM halo, and the parameters $(n, \alpha, \beta, \gamma)$ control the slopes of the density profile in the spike ($r \simeq 2r_s \ll a$), near ($r_s \ll r \ll a$),
far  ($r \gg a$) and transition ($r \simeq a$) regions, respectively. Observations found that $ 0 \le \gamma < 3$ \cite{Zhao96,NFW97}, and $0.005 {\text{kpc}} \lesssim a  \lesssim 50  {\text{kpc}}$, where $\gamma = 0$ corresponds to core models.
Recall that $r_s \equiv 2M = 9.608 \times 10^{-11}(M /10^6\; M_{\odot})\; {\text{kpc}}$. Therefore, we generally have $a \gg r_s$. When $M = 0$,
the above models include several important cases \cite{Zhao96}, such as  $(n, \alpha, \beta, \gamma) = \{(0,1, 4, 1),
(0,1, 3, 1)\}$, which correspond to the Hernquist   \cite{Hernquist90} and   Navarro-Frenk-White (NFW) \cite{NFW97}  profiles.    As mentioned above, when a central BH is present, the trajectories of the DM particles become unstable for $ r < 2r_s$ and will eventually fall inside the BH, whereby a spike is usually developed and the slope $n$ of the spike is generally  model-dependent
\cite{GS99,SFW13,SABB22}. Therefore, we must have
\bqn
\lb{eq2.a}
f = f_o\left(1 - \frac{r_s}{r}\right), \; m =  M, \; \rho = P = 0,  \; (r \le 2r_s),
\eqn where $f_o$ is a constant determined by the matching conditions across  $r  = 2r_s$.
From (\ref{eq1}) we can see that the Einstein field equations contain only the first-order derivatives of $f(r)$ and $m(r)$. Therefore, provided that they are continuous across  $r  = 2r_s$, these equations will be satisfied. For the choice (\ref{eq2}), it is clear that $\rho(r)$ and $P(r)$ are already continuous, while
$f(r)$ can always be made continuous by properly choosing $f_o$.  In the rest of this Letter, we assume that this is always the case without further mentioning it. Thus, we only need to impose the boundary condition $m(2r_s) =  M$.
With such choices of $f_o$ and $m(r)$, the Einstein field equations (\ref{eq1}) are solved in the entire spacetime $r \in (0, \infty)$. In addition,  { in this Letter we consider only the cases in which} the spacetime is asymptotically flat
\bq
\lb{eq2.b}
f(r)  \simeq 1, \quad \frac{m(r)}{r} \simeq 0, \; (r \rightarrow \infty).
\eq

\section{
Analytical Models
}

With the above setup, we are ready to integrate (\ref{eq1}) for the density profile (\ref{eq2}), for which we find a large class of analytical solutions.

{\bf Model I:} This corresponds to the  choice  $(n,\alpha, \beta, \gamma) = (1,1, 4, 1)$ and the mass function is given by
\bqn
\lb{eq3}
m(r) &=& M  + M_h \frac{(r - 2r_s)^2}{(r + a)^2},
\eqn where $M_h \equiv 2\pi\rho_0 a^4/(a+2r_s)$.  Note that $m(\infty) = M + M_h$. Thus, $M_h$ is the total mass of the DM halo.    Observations show that $M  \ll M_h \ll a$ \cite{NFW97,BHS05,SABB22}. In this Letter, we consider the case $5r_s  < M_h < a/10$.
Then, we find
\bqn
\lb{eq3a}
\frac{f'(r)}{f(r)} &=& - \frac{1}{x-a} + \frac{x^2}{{\cal{D}}_1(x)},
\eqn with $x \equiv r + a$, and  {$ {\cal{D}}_1(x) $ is defined in  (\ref{A.1}).
To integrate  (\ref{eq3a}), we first note that  ${\cal{D}}_1(x)  = 0$ has three roots, say,  $x_{1, 2, 3}$, given explicitly in  (\ref{A.2}).  It can be shown that $\Delta  > 0$ defined in  (\ref{A.2a}) for $5r_s  < M_h < a/10$, so that the two roots $x_{2, 3}$  are complex conjugates, $x_3 = x_2^*$, while $x_1$ is real.  Therefore, ${\cal{D}}_1(x)$ can be written as
\bq
\lb{eq3b}
{\cal{D}}_1(x) = \Pi_{n=1}^{3}(x- x_n).
\eq
Substituting this expression into  (\ref{eq3a}) and then integrating it, we find that}
\bqn
\lb{eq3c}
f(r) &=&  \frac{ \Delta_1^{\alpha_1} \Delta_2^{(1-\alpha_1)/2}}{r}\left(\frac{\Delta_1(r)}{\Delta_1}\right) ^{\alpha_1}\left(\frac{\Delta_2(r)}{\Delta_2}\right)^{{(1-\alpha_1)}/{2}} \nb\\
		&& \times
		\exp \left[D_o\left(\tan^{-1}\left(\frac{a+r+k}{\Delta_3}\right)- \frac{\pi}{2}\right)\right], ~~~
\eqn where   {$\alpha_n,\; \Delta_n,\;  \Delta_n(r), \; D_o$ and $k$ are given in   (\ref{A.2b}).
Note that, in writing the above expression for $f(r)$, we chose the integration constant so that the asymptotically flat condition $f(r) = 1$ as $r \rightarrow \infty$   is satisfied [cf.  (\ref{eq2.b})]. With the choice $f_o = 2f(2r_s)$, where $f_o$ is the constant  appearing in  (\ref{eq2.a}),  the continuous condition of $f(r)$ across the surface $r = 2r_s$ is also satisfied.}

By defining ${\cal{G}}(r) \equiv r - 2m(r)$, we find   ${\cal{G}}'(r) = 1 - 4M_h(r -  2r_s)(a + 2r_s)/(r+a)^3 > 0$. Therefore, ${\cal{G}}(r)$  is a monotonically increasing function with a minimal value $r_s$ given at $r = 2r_s$. Since $\rho + P = (1 + m/2{\cal{G}})\rho$,  $\rho + 2P = (1 + m/{\cal{G}})\rho$, we can see that the conditions $\rho \ge 0, \; \rho + P \ge 0$ and $\rho + 2P \ge 0$ always hold for $r \ge 2r_s$, so the DM halo satisfies the weak and strong energy conditions \cite{Hawking:1973uf}.  On the other hand, we also have $\rho - P = \rho[5 - r/{\cal{G}}(r)]/4$. Since ${\cal{G}}(r) \ge r_s$, we find that $m/r = r_s/2r + M_h(r - 2r_s)^2/[r(r+a)^2]
< 1/4 + M_hr/(r+a)^2$, where $r/(r+a)^2$ has its maximal value at $r = a$, so  $m/r < (a+M_h)/4a$ and $r/{\cal{G}}(r) < \underline{}2a/(a-M_h)$. Hence, we have
$\rho - P > (3a - 5M_h)\rho/[4(a-M_h)] > 0$ for $5r_s  < M_h < a/10$. Thus, the halo also satisfies the dominant energy condition \cite{Hawking:1973uf}.

{\bf Model II:} In this model, we choose $(n, \alpha, \beta, \gamma) = (1, 1, 5, 1)$. Then, we find
\bqn
\lb{eq5}
m(r) = M  + M_h\left(1+ 2\frac{(a + 2r_s)^3}{(r + a)^3} -  3\frac{(a + 2r_s)^2}{(r + a)^2} \right),~
\eqn but now with $M_h \equiv 2\pi a^5  \rho_0/[3(a+2r_s)^2]$. Inserting the above expression into (\ref{eq1}), we find that
\bqn
\lb{eq5a}
\frac{f'(r)}{f(r)} =  -\frac{1}{r}  + \frac{y^3}{a{\cal{D}}_{2}(y)},
\eqn where $y \equiv 1 + r/a$, and  ${\cal{D}}_{2}(y)$ is given by  (\ref{A.3a}).
Then, it can be shown that ${\cal{D}}_{2}(y) = 0$ has four roots, two of which are real, denoted by $y_1$ and $y_2$, and the other two are complex conjugates, $y_4^* = y_3 \equiv y_{3R} + i y_{3I}$, all given in (\ref{A.4}).   {Hence, we can follow what we did in the last case, i.e., first write ${\cal{D}}_{2}(y)$
in the form (\ref{eq3b}) but now with four roots, then substitute it into  (\ref{eq5a}),  and finally integrate such obtained $f'(r)/f(r)$, }
we obtain
\bqn
\lb{eq6}
f(r) &=& \frac{\left(y - y_1\right)^{\beta_1}\left(y - y_2\right)^{\beta_2}}{y-1} \left[\left(y- y_{3R}\right)^2 + {y_{3I}}^2\right]^{\beta_3}\nb\\
&& \times
\exp\left\{\hat{\cal{D}}_o\left[\tan^{-1}\left(\frac{y - y_{3R}}{y_{3I}}\right) - \frac{\pi}{2}\right]\right\},
\eqn where $\beta_n$ and $\hat{\cal{D}}_o$ are defined in  (\ref{A.5a}).
Note that the above solution satisfies the asymptotically flat and matching conditions mentioned above.

To show that all three energy conditions also hold in the current case, we first note that ${\cal{G}}''(r) = 12M_h(a+2r_s)^2(3r - a-8r_s)/(a+r)^5$, which vanishes at $r_{\text{min}} = (a+8r_s)/3$. This corresponds to a minimum of ${\cal{G}}'(r)$, where ${\cal{G}}'(r_{\text{min}}) = 1 - 3^4M_h/(4^3(a+2r_s)) > 0$ for $5r_s < M_h < a/10$. Therefore, ${\cal{G}}'(r) > 0$ for
$r > 2r_s$. Hence, in this region ${\cal{G}}(r)$ is monotonically increasing and ${\cal{G}}(r) \ge r_s$. Then, the DM halo satisfies the weak and strong energy conditions. In addition,  the maximum of $[m(r)/r]$ is 1/4 for $5r_s < M_h < a/10$. Then, we find that $\rho - P = (5 - r/{\cal{G}})\rho/4  > (5 - 1/(1 -1/2))\rho > 0$. Hence, the dominant energy condition also holds.

{\bf Model III:} This corresponds to the choice $(n, \alpha, \beta, \gamma) = (1, 1, \frac{5}{2}, 1)$. Then, we find
\bqn
\lb{eq7}
m(r) &=& M + {\cal{M}} \left[\left(1+ \frac{r}{a}\right)^{1/2} - 2 \left(1+\frac{2r_s}{a}\right)^{1/2}\right.\nb\\
&& \left.+ \left(1+\frac{2r_s}{a}\right)\left(1+ \frac{r}{a}\right)^{-1/2}\right],
\eqn where ${\cal{M}} \equiv 8\pi  a^3\rho_0$, and
\bqn
\lb{eq7a}
\frac{f'(r)}{f(r)} &=& - \frac{1}{a(z^2-1)} + \frac{z}{a{\cal{D}}_{3}(z)},
\eqn with $z \equiv (1+r/a)^{1/2}$ and ${\cal{D}}_{3}(z)$ is given in  (\ref{A.6}). Thus,  ${\cal{D}}_{3}(z) = 0$ has three roots, denoted by
$z_{1, 2, 3}$ and given  by (\ref{A.7}). Depending on the values of $u$ and $v$, the nature of these roots is different. In particular, if $u < 0.1709$,  all of them are  real.  If $u > 0.1716$, two of them are complex conjugates, and only one is real, i.e., $z_1$.
If $0.1709 < u < 0.1716$, whether the three roots are all real or not depends on the values of $v$. In the following, we consider the two cases separately.

Before doing so, note that $m(r) \propto \sqrt{r} \rightarrow \infty$ as ${r} \rightarrow \infty$. Thus, the mass of the DM halo becomes infinitely large.
In fact, this is always the case for $\beta \le 3$, including the  NFW model \cite{NFW97}, for which  $\beta = 3$ and $m(r) \propto \ln r$. In reality, a halo is always restricted to a finite region; therefore, one needs to restrict the expression of (\ref{eq7}) to be valid only up to a maximal radius, say, $r_{\infty}$. Then, the total mass of the halo is finite. It is interesting to note that in the current case, the mass does not increase fast enough, so the spacetime is still asymptotically flat, as the conditions (\ref{eq2.b}) still hold.

With the above in mind, let us first consider the case in which all three roots are real. Hence,   {we first write ${\cal{D}}_{3}(z)$ in the form (\ref{eq3b}), then integrate (\ref{eq7a}),  so that finally   we obtain }
\bqn
\lb{eq9}
f(r)=  \frac{a}{r} \frac{\left(z-z_1\right)^{\gamma_1}\left(z-z_2\right)^{\gamma_2}}{\left(z-z_3\right)^{\gamma_1+\gamma_2 -2}},
\eqn where  $\gamma_n$ are given in  (\ref{A.7a}).  The above expressions show that we have $f(\infty) = 1$. On the other hand, from (\ref{eq7}) we find that $m(2r_s) = M$. Therefore,  choosing   $f_o = 2f(2r_s)$, the solution can be smoothly matched to that of the Schwarzschild vacuum solution valid in the region $r \le 2r_s$. We also found that the corresponding DM halo satisfies all three energy conditions. However, the proof is rather tedious and will be presented in detail in \cite{Shen23}. Therefore, here we shall omit the proof. The same will be done for the case in which only one of the three roots is real.

When only one of the three roots is real, which is chosen as $z_1$, we find
\bqn
\lb{eq10}
f(r)&=&\frac{a}{r}\big(z-z_1\big)^{\delta_1}\big((z-z_{2R})^2 + z_{2I}^2\big)^{\frac{2 - \delta_1}{2}}\nb\\
&&\times\exp\left\{\frac{\tilde{\cal{D}}_o}{z_{2I}}\left[\tan^{-1}\left(\frac{z - z_{2R}}{z_{2I}}\right) - \frac{\pi}{2}\right]\right\}, ~~~
\eqn where $\delta_n$ are given in  (\ref{A.7b}).
Thus, we have $f(\infty) = 1$ and $m(r)/r \rightarrow 0$ as $r \rightarrow \infty$.
As a result, spacetime is asymptotically flat, despite the fact $m(r) \sim {\cal{M}} \sqrt{r/a}$ for $r \gg a$.

{\bf Models IV $\&$ V:} These two models have the same choice of $(\alpha, \beta, \gamma) = (1, 4, 0)$ but with $n = 1, 2$, respectively. In contrast to the above three models, the density profiles now become constants when $2r_s \ll r \ll a$, which are often referred to as the core models. Then, the mass functions are given by
\bqn
\lb{eq11}
m^{\text{(N)}}(r) &=& M +  M_{h}^{\text{(N)}} \nb\\
&& + \sum_{J = 1}^{3}\frac{{\cal{A}}^{\text{(N)}}_J}{(1+r/a)^J}, \; ({\text{N}} = 4, 5), ~~
\eqn where  $M_{h}^{\text{(N)}} $ and ${\cal{A}}^{\text{(N)}}$ are given in (\ref{eq11a}).
Hence, we find
\bqn
\lb{eq12}
\frac{f'(r)}{f(r)}  &=& -  \frac{1}{aw} + \frac{(1+w)^3}{{\cal{D}}^{\text{(N)}}(w)},
\eqn where  $w \equiv r/a$ and ${\cal{D}}^{\text{(N)}}(w)$ is defined in (\ref{eq12a}).
It can be shown that in both cases
${\cal{D}}^{\text{(N)}}(w) = 0$ has four roots given in (\ref{A.8}), two of which are real and the other two are complex conjugates, denoted respectively by $w^{\text{(N)}}_{j (= 1, ..., 4)} $, where $w^{\text{(N)}}_3 = w^{\text{(N)}}_{3R} + i w^{\text{(N)}}_{3I}$ and $w^{\text{(N)}}_3$ and  $w^{\text{(N)}}_4$ are complex conjugates.
Then,  {we have write ${\cal{D}}^{\text{(N)}}(w)$ in the form (\ref{eq3b}), with which the integration of  (\ref{eq12}) yields}
\bqn
\lb{eq13}
f(r) &=& \frac{(r - a w_1)^{\epsilon_1}(r - aw_2)^{\epsilon_2}}{a^{\epsilon_1 + \epsilon_2 - 1}r} \left[\frac{(r-aw_{3R})^2}{a^2}+w_{3I}^2\right]^{\epsilon_3}\nb\\
&\times& \exp\left\{\frac{\epsilon_3 w_{3R}+\epsilon_4}{w_{3I}}\left[\tan^{-1}\left(\frac{\frac{r}{a}-w_{3R}}{w_{3I}}\right)-\frac{\pi}{2}\right]\right\},\nb\\
\eqn where $\epsilon_N$ are given in (\ref{eq14}).
Note that we have dropped the superscript ``(N)" in writing down  (\ref{eq13}) and (\ref{eq14}) without causing any confusion.
In addition, we set the integration constant so that the asymptotically flat conditions (\ref{eq2.b}) and the matching conditions across $r = 2r_s$ are all satisfied.  In these two core models, the DM halo satisfies the three energy conditions \cite{Shen23}.

\section{
Conclusions
}

The construction of these analytic models is expected to significantly facilitate the studies of QNMs of SMBHs and GWs emitted by IMRIs and EMRIs, which are the ideal sources of next-generation GW detectors, both ground- and space-based  detectors \cite{ET,CE,LISA,TianQin,Taiji,DECIGO}.  As the small compact companion of the SMBH passes through the DM halo, it feels gravitational dragging forces, whereby its orbits and the gravitational waveforms emitted by it are modified. The modifications depend on the nature and properties of the DM halo. Such modifications can be detected by LISA and other next generation detectors \cite{Eda15,Kavanagh20,Coogan22,Hannuksela20,Zi21,Fan22,Berezhiani23}. Hence, it opens a brand new window to studying DM.

Certainly, the applications of these models are not only restricted to the studies mentioned above but also equally applicable to other fields, such as shadows and lensing of SMBHs, which can be the sources of EHT observations \cite{EHT22,EHT19},
the dynamics of galaxy systems, including our own Milky Way galaxy \cite{BHS05,Freese09,SBD17,WT18,AM21,Hernquist90,NFW97}, and a better understanding of the cusp-core problem \cite{MUL23},
To name a few.

Note that for EHT observations, baryons are often required to be present.  In the current models, we consider DM as the dominant matter component and consider baryons (as well as other matter fields) as perturbations. Therefore, to the leading order, they are assumed to be negligible.
In our forthcoming investigations, we hope to include such effects.

Finally, we note that using the generalized Newman-Janis algorithm \cite{AA14}, we can construct rotating SMBHs surrounded by DM halos. In the Boyer-Lindquist coordinates, they are given precisely by (16) and (17) of \cite{Liu20}.

\section*{acknowledgments}

We thank Profs. Katherine Freese and Monica Valluri for their valuable comments and suggestions. Z.S. thanks the hospitality of the Physics Department and Baylor University during his visit.
A.W. is partially supported  by the  U.S. Natural Science Foundation (NSF) under Grant No.  PHY2308845. Y.G. is partially supported by the National Key Research and Development Program of China under Grant No. 2020YFC2201504.

\section*{Appendix A: Roots  for Models I - V}
\renewcommand{\theequation}{A.\arabic{equation}}
\setcounter{equation}{0}

{\bf Model I:} In this model, the function ${\cal{D}}_1(x)$ appearing in  (\ref{eq3a}) is defined as
\begin{widetext}
\bqn
\lb{A.1}
{\cal{D}}_1(x) \equiv x^3 -Ax^2 +Bx - C,\quad
A   \equiv  a + r_s   + 2M_h, \quad B \equiv 8\pi a^4\rho_0, \quad  C \equiv    2M_h(a+2r_s)^2.
\eqn
Then,  $ {\cal{D}}_1(x) = 0$
has the following three roots
	\bqn
	\lb{A.2}
	x_1&=&\sqrt[3]{\frac{1}{2}\left(-q+\sqrt{\Delta}\right)}+\sqrt[3]{\frac{1}{2}\left(-q-\sqrt{\Delta}\right)}{+\frac{A}{3}},\nb\\
	x_2&=&\frac{-1+\sqrt{3}i}{2}\sqrt[3]{\frac{1}{2}\left(-q+\sqrt{\Delta}\right)} - \frac{1+\sqrt{3}i}{2}\sqrt[3]{\frac{1}{2}\left(-q-\sqrt{\Delta}\right)}{+\frac{A}{3}}, \quad x_3 = x_2^*,
	\eqn where  $x \equiv r+a$, and
\bqn
\lb{A.2a}
	p \equiv B-\frac{1}{3} {A^2},  \quad q \equiv - \frac{2}{27} A^3 + \frac{1}{3} {AB}- C,\quad
\Delta       \equiv  q^2+\frac{4}{27} p^3.
\eqn
In addition, the quantities, $\Delta_n(r), \; \Delta_n, \; k$ and $\alpha_n$ appearing in  (\ref{eq3c}) are defined as follows:
\bqn
\lb{A.2b}
\Delta_1(r) &\equiv& a+r-x_1,  \quad
\Delta_2(r) \equiv (a+r+k)^2+|x_2|^2{-k^2}, \quad
\Delta_{1, 2} \equiv \Delta_{1, 2}(2r_s), \quad \Delta_3 \equiv \sqrt{|x_2|^2-k^2}, \nb\\
D_o &\equiv& \frac{1}{{\Delta_3}} \left({{\alpha_2}-k (1-\alpha_1)}\right), \quad
k \equiv \frac{1}{2}\left(\sqrt[3]{\frac{1}{2}\left(-q+\sqrt{\Delta}\right)}+
	\sqrt[3]{\frac{1}{2}\left(-q-\sqrt{\Delta}\right)}\right){-\frac{A}{3}},  \nb\\
	\alpha_1 &\equiv& \frac{x_1^2}{x_1^2+|x_2|^2+2k x_1}, \;\;\;
	\alpha_2 \equiv \frac{|x_2|^2}{x_1}\alpha_1.
\eqn

{\bf Model II:} In this model,  the function ${\cal{D}}_{2}(y)$ appearing in  (\ref{eq5a}) is defined as
\bqn
\lb{A.3a}
{\cal{D}}_{2}(y) &\equiv& y^4 - {\cal{A}} y^3 + {\cal{B}} y - {\cal{C}},\quad
{\cal{A}} \equiv 1 + \frac{2}{a}\left(M+M_h\right), \quad
{\cal{B}} \equiv \frac{6}M_h}{a^2\left(a+2r_s\right)^2,\quad
{\cal{C}} \equiv \frac{4M_h}{a^4}\left(a+2r_s\right)^3.
\eqn
Then, the equation $ {\cal{D}}_{2}(y) = 0$ has four roots
	\bqn
	\lb{A.4}
	y_1&=&-\frac{1}{4}\left(b-\sqrt{b^2+4y_0}\right)+\frac{1}{4}\sqrt{\left(b-\sqrt{b^2+4y_0}\right)^2-8\left(y_0-\sqrt{y_0^2-4c}\right)},\nb\\
	y_2&=&-\frac{1}{4}\left(b-\sqrt{b^2+4y_0}\right)-\frac{1}{4}\sqrt{\left(b-\sqrt{b^2+4y_0}\right)^2-8\left(y_0-\sqrt{y_0^2-4c}\right)},\nb\\
	y_3&=&-\frac{1}{4}\left(b+\sqrt{b^2+4y_0}\right)+\frac{1}{4}\sqrt{\left(b+\sqrt{b^2+4y_0}\right)^2-8\left(y_0+\sqrt{y_0^2-4c}\right)},\nb\\
	y_4&=&-\frac{1}{4}\left(b+\sqrt{b^2+4y_0}\right)-\frac{1}{4}\sqrt{\left(b+\sqrt{b^2+4y_0}\right)^2-8\left(y_0+\sqrt{y_0^2-4c}\right)},
	\eqn where
\bqn
\lb{A.5}
b&=&-(1+2u+2v), \quad
c = -4u(1+4v)^3,\quad
y_0 = \left(-\frac{q}{2}+\sqrt{\frac{q^2}{4}+\frac{p^3}{27}}\right)^{\frac{1}{3}}-\left(\frac{q}{2}+\sqrt{\frac{q^2}{4}+\frac{p^3}{27}}\right)^{\frac{1}{3}},\nb\\
p&=&-6u(1 + 4v)^2(1 + 2 u + 2 v) + 16 u(1 + 4v)^3,\quad
q = \quad4 u (1 +4 v)^3 (1 + 2 u + 2 v)^2 - 36 u^2 (1 + 4v)^4,
\eqn and $u \equiv {2\pi a^4\rho_0}/[{3(a+ 4M)^2}],\; v \equiv {M}/{a}$.
In addition, the coefficients $\beta_n,\; \hat{\cal{D}}_o$ and $D_n$ appearing in  (\ref{eq6}) are defined by
\bqn
\lb{A.5a}
\beta_1 &\equiv& \frac{y_1^3}{(y_1-y_2)D_1}, \quad \beta_2 \equiv \frac{y_2^3}{(y_2-y_1) D_2},\quad
\beta_3 \equiv \frac{1}{2}(1 - \beta_1 - \beta_2), \quad
\beta_4 \equiv \frac{|y_3|^2\left[\left(y_1+y_2\right)y_3^2-2y_{3R}y_1y_2\right]}{D_1D_2},  \nb\\
D_n &\equiv& |y_3|^2 + y_n^2 - 2 y_{3R} y_n, \quad
\hat{\cal{D}}_o \equiv \frac{\beta_4 + 2y_{3R}\beta_3}{y_{3I}}.
\eqn

{\bf Model III:} In this model, the function $ {\cal{D}}_{3}(z)$ appearing in  (\ref{eq7a}) is defined by
\bqn
\lb{A.6}
{\cal{D}}_{3}(z) &\equiv& z^3-uz^2+ \left(2u\sqrt{1+4v}-2v-1\right)z
u(1+4v), ~~~~~
\eqn with $z \equiv (1+r/a)^{1/2}$,  $v \equiv M/a$ and $ u \equiv 2{\cal{M}}/a$.
Then, the equation ${\cal{D}}_{3}(z) = 0$
has three roots $z_{1, 2, 3}$, given, respectively, by
	\bqn
	\lb{A.7}
	z_1&=&\sqrt[3]{\frac{1}{2}\left(-q+\sqrt{\Delta}\right)}+\sqrt[3]{\frac{1}{2}\left(-q-\sqrt{\Delta}\right)}+\frac{u}{3},\nb\\
	z_2&=&\frac{-1+\sqrt{3}i}{2}\sqrt[3]{\frac{1}{2}\left(-q+\sqrt{\Delta}\right)} - \frac{1+\sqrt{3}i}{2}\sqrt[3]{\frac{1}{2}\left(-q-\sqrt{\Delta}\right)}+\frac{u}{3},\nb\\
	z_3&=&\frac{-1-\sqrt{3}i}{2}\sqrt[3]{\frac{1}{2}\left(-q+\sqrt{\Delta}\right)} - \frac{1-\sqrt{3}i}{2}\sqrt[3]{\frac{1}{2}\left(-q-\sqrt{\Delta}\right)}+\frac{u}{3}.
	\eqn
In addition, the constants $\gamma_n$ appearing in  (\ref{eq9}) fir three real roots are given by
\bqn
\lb{A.7a}
\gamma_1 &\equiv& \frac{2z_1^2(z_2-z_3)D_4}{z_3(z_1-z_2)}, \quad
\gamma_2 \equiv \frac{2z_2^2(z_1-z_3)D_4}{z_3(z_2-z_1)}, \quad
D_3 \equiv z_1(z_2^2-z_3^2)+z_2(z_3^2-z_1^2)+z_3(z_1^2-z_2^2),\nb\\
D_4 &\equiv& \frac{z_3(z_2-z_1)}{D_3}.
\eqn with $z_{1,2,3}$ being the three real roots given in (\ref{A.7}).

When only one of the three roots is real, the integration of  (\ref{eq7a}) yields the solution of $f(r)$ given by  (\ref{eq10}) with
\bqn
\lb{A.7b}
\delta_1 &=& \frac{2az_1^2}{\hat{\cal{D}}_3}, \quad
z_{2} = z_{2R} + i z_{2I}, \quad
\tilde{\cal{D}}_o =  2a(K_3 + 2z_{2R}K_2)z_{2R} - 2a|z_2|^2K_2, \quad
K_2 = - \frac{z_1}{\hat{\cal{D}}_3}, \quad
K_3 = \frac{|z_2|^2}{\hat{\cal{D}}_3}, \nb\\
\hat{\cal{D}}_3 &\equiv& a[z_1^2 + |z_2|^2 - 2z_1 z_{2R}].
\eqn

{\bf Models IV $\&$ V:}  The coefficients appearing in (\ref{eq11}) are given by
\bqn
\lb{eq11a}
&&  M_h^{(4)} = \frac{4\pi a^3\rho_0}{3(1+2r_s/a)^{2}} \left(1+\frac{3r_s}{a}\right), \quad
{\cal{A}}^{(4)}_1 = -\frac{3M_h^{(4)}}{1+3r_s/a} \left(1+\frac{2r_s}{a}\right)^2, \quad
{\cal{A}}^{(4)}_2 = \frac{3M_h^{(4)}}{1+3r_s/a} \left(1+\frac{2r_s}{a}\right)^2\left(1+\frac{r_s}{a}\right),\nb\\
&&   {\cal{A}}^{(4)}_3 = -\frac{M_h^{(4)}}{1+3r_s/a} \left(1+\frac{2r_s}{a}\right)^3, \quad  M_h^{(5)}  = \frac{4\pi a^3\rho_0}{3(1+2r_s/a)}, \nb\\
&&
{\cal{A}}^{{(5)}}_1 = -3M_h^{(5)}\left(1+\frac{2r_s}{a}\right),\quad
{\cal{A}}^{{(5)}}_2 =  3 M_h^{(5)}\left(1+\frac{2r_s}{a}\right)^2,\quad
{\cal{A}}^{(5)}_3 =  -M_h^{(5)}\left(1+\frac{2r_s}{a}\right)^3.  ~~~
\eqn
The function ${\cal{D}}^{\text{(N)}}(w)$   of (\ref{eq12})  is defined as
\bqn
\lb{eq12a}
{\cal{D}}^{\text{(N)}}(w) &\equiv& {\cal{D}}_0^{\text{(N)}} + \left(a-2{\cal{A}}^{\text{(N)}}_2-{\cal{A}}^{\text{(N)}}_1-6{\cal{A}}^{\text{(N)}}_0\right)w
+\left(3a-2{\cal{A}}^{\text{(N)}}_1-6{\cal{A}}^{\text{(N)}}_0\right)w^2 \nb\\
&& +  \left(3a-2{\cal{A}}^{\text{(N)}}_0\right)w^3 + aw^4,\; (w \equiv {r}/{a}), ~~~~~~
\eqn with ${\cal{D}}_0^{\text{(N)}} \equiv -2\sum_{n=0}^{3}{{\cal{A}}^{\text{(N)}}_n}$, ${\cal{A}}^{\text{(N)}}_0 \equiv M + M_h^{\text{(N)}}$.
The coefficients of (\ref{eq13})  are given by
\bqn
\lb{eq14}
\epsilon_1&=&\frac{(1+w_1)^3}{(w_1-w_2)d_1}, \;\;\;
\epsilon_2 = \frac{(1+w_2)^3}{(w_2-w_1)d_2},\quad \epsilon_3 \equiv (1-\epsilon_1 - \epsilon_2)/2, \quad d_n \equiv w_n^2+|w_3|^2-2w_{3R}, \nb\\
\epsilon_4&=&\frac{1}{d_1d_2}\bigg\{4w_{3R}^2\left(3+w_1+w_2\right)
+w_1\left(w_2-3|w_3|^2  -3w_2|w_3|^2+|w_3|^4\right)
+2w_{3R}\left[1+3w_2-6|w_3|^2\right.\nb\\
&&\left. -2w_2|w_3|^2+w_1\left(3+3w_2-2|w_3|^2\right)\right]  + |w_3|^2\left(-1+3|w_3|^2-3w_2+w_2|w_3|^2\right)\bigg\},
\eqn
Then,   $ {\cal{D}}^{\text{(N)}}(w)  = 0$ has four roots, $w_{1, ..., 4}$, given respectively by
\bqn
\lb{A.8}
w_1&=&\frac{1}{2}\left(-B_1 +\sqrt{B_1^2-4C_1}\right), \quad w_2 = \frac{1}{2}\left(-B_1 - \sqrt{B_1^2-4C_1}\right), \nb\\
w_3 &=&\frac{1}{2}\left(-B_2 +i\sqrt{4C_2- B_2^2}\right),\quad    w_4 = \frac{1}{2}\left(-B_2 -i\sqrt{4C_2- B_2^2}\right),
\eqn where
\bqn
\lb{A.9}
B_1&=&\frac{1}{2}\left(b-\sqrt{b^2-4c+4y_0}\right), \quad  B_2 = \frac{1}{2}\left(b+\sqrt{b^2-4c+4y_0}\right),\nb\\
C_1 &=&\frac{1}{2}\left(y_0-\sqrt{y_0^2-4e}\right),  \quad C_2 = \frac{1}{2}\left(y_0+\sqrt{y_0^2-4e}\right),\nb\\
y_0&=&\sqrt[3]{-\frac{q}{2}+\sqrt{\left(\frac{q}{2}\right)^2+\left(\frac{p}{3}\right)^3}} + \sqrt[3]{-\frac{q}{2}-\sqrt{\left(\frac{q}{2}\right)^2+\left(\frac{p}{3}\right)^3}} +\frac{c}{3}, \nb\\
p &=&bd-4e-\frac{c^2}{3},\quad
q = -b^2e+4ce-d^2 + \frac{c}{3}\left(bd-4e\right)-\frac{2}{27}c^3,\nb\\
b&=&3-\frac{2\left(M + M_h\right)}{a}, \quad
c = 3-\frac{1}{a}\left[2{\cal{A}}_1+6\left(M + M_h\right)\right],\nb\\
d&=&1-\frac{2}{a}\left[{\cal{A}}_2+2{\cal{A}}_1+3\left(M + M_h\right)\right], \quad
e = -\frac{2}{a}\left[{\cal{A}}_3+{\cal{A}}_2+{\cal{A}}_1+\left(M + M_h\right)\right].
\eqn
\end{widetext}
Note that in writing (\ref{A.8})-(\ref{A.9}), we have dropped the super-indices ``(N)" without causing any confusion.

\end{document}